# Electronic Structure of Bimetallic CoRu Catalysts Modulates SWCNT Nucleation

Alister J. Page*,[1] Dan Villamanca,[1] Placidus B. Amama,[2] Ben McLean,[3]

[1]Discipline of Chemistry, College of Engineering, Science and Environment, The University of Newcastle, Callaghan NSW 2308, Australia

[2] Tim Taylor Department of Chemical Engineering, Kansas State University, Manhattan, KS 66506, USA

[3]School of Engineering, RMIT University, Victoria, 3001, Australia

*Corresponding Author.  Email: alister.page@newcastle.edu.au

**Abstract**

Nucleation of single-walled carbon nanotubes (SWCNTs) via chemical vapour deposition of methane on CoRu bimetallic nanoparticles is simulated using quantum chemical molecular dynamics. By varying the Ru loading in the catalyst, we show that Ru decreases catalytic efficiency; C-H bond activation is impeded, key reactive intermediate species become longer-lived on the catalyst surface, and longer carbon chains are stabilised through the earliest stages of SWCNT nucleation. Analysis of the CoRu nanoparticle structure during the CVD process shows that this influence of Ru is indirect, with the catalyst adopting Ru-Co core-shell or segregated structures throughout nucleation, and Co exclusively driving the catalytic decomposition of the methane precursor. We show that the influence of Ru occurs via the electronic structure of the catalyst itself, by lowering the Fermi level of the catalyst due to lower energy 4d/5s states, in a manner consistent with d-band theory.

**Introduction**

Since their discovery in the early 1990s,[1] single-walled carbon nanotubes (SWCNTs) have been studied extensively due to their remarkable optoelectronic[2] mechanical properties.[3,4] During this time, synthetic approaches have been tuned and optimised for prospective applications, such as energy, microelectronic and composite material technologies.[5] Simultaneously, our fundamental understanding of how SWCNTs nucleate and grow has been refined,[6–13] as has our ability to control these processes.

The properties of SWCNTs are intrinsically linked to their atomic scale helicity, i.e. 'chirality', and intense effort has been dedicated to realising so-called 'chirality-controlled' SWCNT growth over the last three decades. Catalyst design approaches, based on the premise that SWCNT chirality control can be achieved via the composition and structure of the catalyst, has been a prominent avenue of research during this time. In this respect Co-based catalysts have been studied extensively. In many cases, SWCNT chirality and diameter selectivity arise from the stabilisation of the Co catalyst against sintering and reduction by the co-catalyst, rather than the formation of any particular alloy or intermetallic catalyst phase. For instance, Co/Mo bimetallic catalysts show selectivity for growing narrow diameter distributions[14] and specifically (6,5) and (7,5) SWCNTs,[15,16] and sub-nanometer diameter distributions[17] due to this effect. Similar results were reported more recently Cui et al.,[18] who showed that Co/Cu bimetallic catalysts selectively produce SWCNTs with sub-nanometer diameters due to Cu 'anchoring' the active Co nanoparticles, preventing their reduction and sintering. Amama et al.[19] have suggested, in the case of CoRu bimetallic catalyst nanooparticles, that improvements in SWCNT quality and selectivity due to inhibited sintering arise from Ru increasing the Co nanoparticle cohesive energy.

A complementary strategy for catalyst design that has emerged more recently is the use of crystalline intermetallic compounds for SWCNT growth. W/Co bimetallic catalysts are notable in this respect, following the report of Yang et al.[20] who demonstrated near-selective growth of (12,6) SWCNTs on solid-phase $W_7Co_6$ nanocrystal catalysts. Chirality selectivity here is attributed to optimal lattice matching between the (0 0 12) facet of the $W_7Co_6$ nanocrystal and the growing (12,6) SWCNT edge. However, An et al.[21,22] subsequently highlighted the importance of an intermediate $Co_6W_6C$ catalyst structure during SWCNT growth. Wang et al.[23] have also studied the SWCNT growth mechanism from Co-W-C nanoparticles, showing that growth is supported exclusively via a single-phase cubic η-carbide catalyst. The complexities intrinsic to understanding SWCNT growth mechanisms on bi- and tri-metallic catalysts was recently demonstrated by Ji et al.[24] using a high-throughput investigation of Co/Mo/Pt ternary catalysts. This work revealed active growth phases including CoPt alloy nanoparticles support the growth of semiconducting SWCNTs, while Janus Co-Mo/Co catalyst phases yield long growth lifetimes and high SWCNT yields. Similar complexity was also demonstrated by Shiina et al.,[25] who reported exclusive growth of (6,5) SWCNTs using a ternary NiSnFe catalyst.

The relationship between bimetallic catalyst composition, structure and SWCNT growth has therefore been studied extensively. However, the impact that bimetallic catalyst composition has on the elementary aspects of the CVD mechanism itself remains largely unknown. Typically such insights are provided by molecular simulations, such as density functional theory (DFT) and molecular dynamics. The principal limitation here is presumably the fact that custom-made empirical[26,27]/machine-learning[28,29] potentials or tight-binding parameters[30,31] are required to avoid the prohibitive computational cost of density functional theory (DFT). Such potentials have provided a deep understanding of SWCNT nucleation and growth over the last three decades. However, the development of these methods is non-trivial, and they are typically non-transferable between catalysts/growth conditions etc.

Herein we address this knowledge gap specifically for the case of CoRu bimetallic catalysts. Motivated by the recent report of Amama et al., we use the GFN1-xTB[32] tight binding method to simulate methane CVD on model CoRu bimetallic catalyst nanoparticles. Under simulated CVD conditions, these catalyst nanoparticles are predominantly Ru-Co phase segregated and Ru-Co core-shell structures, with the catalytic activity restricted exclusively to the Co surface. Our simulations show however that Ru incorporation into Co nanoparticles fundamentally modulates the catalytic activity of the nanoparticle itself. Specifically, increasing the nanoparticle Ru content between 0-30% leads to a decrease in C-H bond activation, more persistent populations of key CVD reactive intermediates, longer and more stable carbon chains and increased ring growth prior to SWCNT nucleation. Analysis of the nanoparticle electronic structure shows that Ru influences CVD surface chemistry indirectly via modulating the Fermi level of the nanoparticle catalyst. In light of these results we propose that established heuristic concepts in heterogeneous catalysis, such as d-band theory,[33,34] can be used to understand fundamental aspects of CVD SWCNT growth on bimetallic catalysts.

## Methods

### Quantum Chemical Methods

Methane CVD and CNT nucleation is simulated here using the GFN1-xTB method of Grimme et al. This method is based on a one-centred self-consistent density functional tight binding (SCC-DFTB) scheme that is similar in some respects to the original SCC-DFTB method[35] used extensively in previous investigations of CVD mechanisms.[36–39] The key difference is that there are no two-centred terms in the GFN1-xTB Hamiltonian, meaning that the pairwise terms (Hamiltonian, overlap and repulsive potential) in SCC-DFTB are avoided. This means that GFN1-xTB can be applied with no further parameterisation to more complex systems, such as alloy catalyst nanoparticles. Considering that the chemical structures and processes taking place during a CVD process are somewhat beyond the training set of the GFN1-xTB method itself, we verify that this method is providing qualitative consistency with the generalised gradient approximation (GGA) level of density functional theory (PBE functional[40]) in supporting information (Figures S1,2). All GFN1-xTB calculations were performed using the DFTB+ software,[41] and PBE calculations employed the Vienna Ab Initio Simulation Package (VASP).[42–45]

### Molecular Dynamics Simulations

The molecular dynamics (MD) equations of motion were integrated using the velocity-Verlet algorithm with a time step of 1 fs. An electronic temperature of 300 K was used throughout, to ensure efficient electronic convergence of the SCC cycles at each MD timestep. A constant NVT ensemble with a temperature of 1,000 K was applied via a Nosé-Hoover chain thermostat[46–48] with a chain-length of 3 and a coupling strength of 1,000 cm$^{-1}$.

CVD is simulated here on $Co_{55-x}Ru_x$ bimetallic catalyst nanoparticles, where $x$ = 0, 6, 11 and 17. These values correspond to Ru loadings of ~0%, 10%, 20% and 30%, and therefore span a range relevant to recent high-throughput experiments studying CNT growth on supported CoRu catalysts.[19] $Co_{55}$ nanoparticle structures were generated from the bulk lattice using the atomic simulation environment,[49] atoms were then randomly selected and replaced with Ru for each individual simulations. CVD simulations were replicated five times for each $Co_{55-x}Ru_x$ catalyst nanoparticle (20 distinct GFN1-xTB/MD simulations in total). All reported quantities in the discussion below are based on average values observed across each set of five trajectories. All $Co_{55-x}Ru_x$ catalyst nanoparticles were equilibrated to the simulation temperature for a period of 10 ps before methane was introduced. Methane deposition was simulated by periodic adsorption of $CH_x$ species at intervals of 2 ps, until a surface carbon density of 60 carbon atoms was achieved.[50] After that point, the system was annealed with a constant surface carbon density at the simulation temperature until a total simulation time of 500 ps was reached. Supporting Movies S1-3 provide examples of an individual trajectory at each Ru loading.

**Nanoparticle Fermi Level Analysis**

Fermi level analysis was performed using an in-house python code, and is based on a random-selection of 1,000 individual frames selected from the five replicate trajectories for each $Co_{55-x}Ru_x$ catalyst nanoparticle CVD simulation. Ultimately this represents a 0.2% sampling density of each $Co_{55-x}Ru_x$ catalyst nanoparticle CVD simulation condition. Once selected, the geometry of the system was pre-processed to remove all C and H atoms, and the electronic structure of the $Co_{55-x}Ru_x$ catalyst nanoparticle structure remaining was calculated using GFN1-xTB (without geometry optimisation).

**Results and Discussion**

**CoRu Catalyst Nanoparticle Structure**

We begin our discussion by considering the influence Ru content has on the physical structure of the CoRu bimetallic nanoparticles in Figure 1. Figure 1a shows the radial distribution functions of Ru, Co and C observed during simulated CVD. Irrespective of the Ru content present, the CoRu nanoparticles resemble something closer to a core-shell structure. The greatest overlap between Co and Ru distributions is observed for the highest Ru loading considered here, 30%, but still there is effectively no presence of Ru at the active catalyst interface in this case. These results do not change appreciably across a temperature range of 300-2100 K (Figure S3). We note however the CoRu nanoparticles' tendency to become more segregated at lower temperatures. In particular, for temperatures up to ~800 K (just below the melting point) we observe a significant proportion of the Co atoms residing preferentially at the nanoparticle centre, followed by Ru, and then Co, with increasing radial distance. There is also evidence of increased crystallinity at the lowest temperatures considered here, as anticipated.

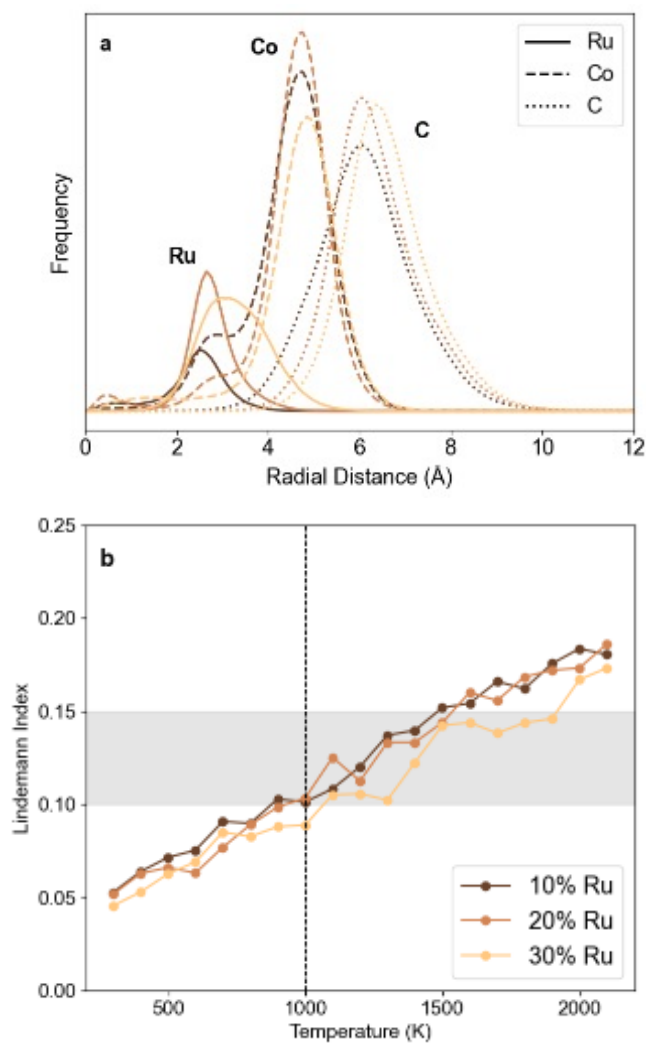

**Figure 1.** (a) Radial distribution functions of Ru and Co in CoRu nanoparticles during simulated methane CVD indicate the catalyst nanoparticles exhibit a core-shell structure, and are not truly alloyed. (b) Average nanoparticle Lindemann indices for CoRu nanoparticles between 300 and 2100 K observed from 50 ps of constant temperature GFN1-xTB/MD simulation (in the absence of carbon). The grey band indicates the typically-accepted range in the Lindemann index that corresponds to the solid-liquid phase transition, the dashed line indicates the simulated CVD

temperature (1000 K). Lindemann indices for Co and Ru components of the CoRu nanoparticles are presented in Figure S4.

The Ru loading only marginally changes the phase behaviour of the CoRu nanoparticles. Figure 1b shows that, across the range of Ru contents considered here, the melting points of the CoRu nanoparticles employed here lie in the range of ~1000-1500 K, typical of transition metal nanoparticles with nanometer-scale diameters.[51] The melting point transition for the 10% and 20% Ru nanoparticles are comparable according to Figure 1b, whereas the 30% Ru nanoparticles melt at a slightly lower temperature. The CVD simulations discussed below were carried out at a temperature of 1000 K, which sits at the lower limit of this melting point range. However, we do not consider the CoRu nanoparticles to be truly solid-phase during our simulated CVD process, considering the individual contributions that Co and Ru atoms make to the average Lindemann index (Figure SX2). The Co component of the CoRu nanoparticle, predominantly at the nanoparticle surface, is molten at temperatures above ~800 K for the range of Ru loadings considered here. By comparison, the average Lindemann index for the Ru component of the CoRu nanoparticles is consistently lower than that of the Co component. This result is consistent with the radial distributions shown in Figure 1a; as Ru atoms in these CoRu nanoparticles more often aggregate in the core, they have a tendency to have a lower average thermal motion. However, the nanoparticles employed here can only be considered truly solid well below temperatures of ~800-900 K, i.e. below the CVD temperature employed.

**Surface Catalysis on CoRu Nanoparticles**

The Ru-Co core-shell structure observed in Figure 1 therefore suggests that the CVD process on these CoRu nanoparticles is largely driven by the catalytically active Co atoms on the nanoparticle surface, while Ru atoms are unlikely to be directly involved in CVD. We examine this hypothesis in Figure 2, by quantifying the key bonding interactions observed in our GFN1-xTB/MD CVD simulations. It is immediate from this figure that the Co atoms in the catalyst nanoparticle are the principal component driving the surface adsorption and catalytic activation of the methane precursor, as evidenced by the populations of Co-H and Co-C bonds (Figure 2a,b). By comparison, Ru-H and Ru-C bond populations (Figure 2c,d) during this period are negligible.

Movies S1-3 show that, interestingly, following carbon adsorption the nanoparticles frequently convert from a core-shell-like structure to a phase-segregated structure. Nevertheless, carbon remains exclusively adsorbed to the Co component, consistent with the relative carbon solubilities of Ru and Co at elevated temperatures.[52] Despite Ru not actively participating in the catalytic decomposition of the methane precursor however, it has a significant impact on the catalytic activity of the nanoparticle overall. For instance, as Ru content increases so does the population of C-H bonds, meaning that C-H bond activation decreases (Figure 2b). A consistent trend is observed for the population of Co-H bonds (Figure 2d), which decrease significantly with increasing Ru content, suggesting that Ru prevents Co from catalytically activating C-H bonds. The influence of Ru on carbon adsorption can be gauged via the population of Co-C bonds. Figure 2b shows that Ru's effect is most noticeable in the period where methane precursors are actively being supplied to the nanoparticle surface (~0-200 ps). In the post-supply annealing period, the number of Co-C bonds for all simulations converges to a value of ~2.0, irrespective of Ru content. The level of surface saturation on the nanoparticle is therefore not influenced strongly by the amount of Ru present in the catalyst. This surface saturation is also evident in Figure 1, which shows a significant overlap between the radial distribution of C and Co atoms, indicating that the immediate subsurface of the nanoparticle is carbon rich. Considering the distribution of Ru atoms, it is likely that this surface/subsurface carbon is responsible for the small number of Ru-C and Ru-H bonds observed Figure 2e,f.

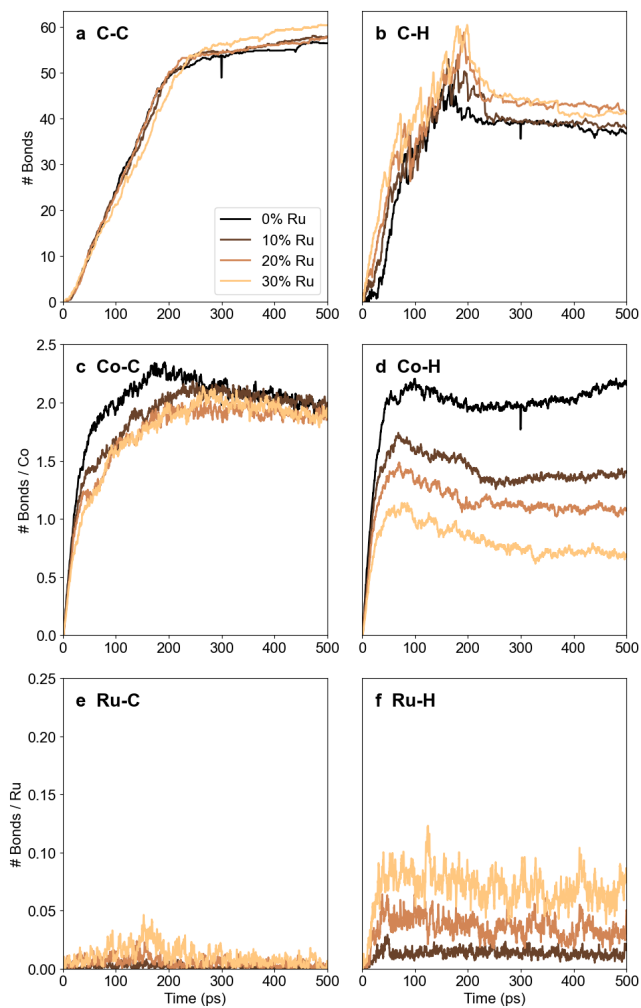

**Figure 2.** Bond populations indicate that Co drives the catalytic process. Evolution of key bonding interactions during simulated methane CVD on CoRu nanoparticles as a function of nanoparticle Ru content, (a) C-C, (b) C-H (c) Co-C (d) Co-H, (e) Ru-C and (f) Ru-H. Data in (c,d) and (e,f) are normalised against the number of Co and Ru atoms in the nanoparticle, respectively. Note the different vertical scales in (a,b), (c,d) and (e,f)

The trends in the [Ru,Co]-[C,H] bond populations shown in Figure 2 are broadly consistent with the population of key CVD precursor species present on the nanoparticle surface during the simulated CVD process. These populations are shown in Figure 3, viz. for activated $C_1$ and $C_2$ species (additional population data is provided in Figure S5). The populations of the key $C_1$ species $CH$, $CH_2$ and $CH_3$ reflects the distributions of the species that are adsorbed on the nanoparticle surface in our CVD algorithm (i.e. $CH_3 > CH_2 > CH$). In each case however, Figure 3a-c show that the active Co surface on the CoRu nanoparticles activate the adsorbed precursor immediately; the populations of each of these species is quenched following the first 200 ps of the simulation, i.e. during which period in the simulation when carbon is being adsorbed to the surface. The mechanism by which carbon chain growth proceeds from these elementary nucleation precursors is well documented in the literature.[6,27,29,53] Notably, a principal driver of chain growth and nucleation is the $C_2H$ radical (Figure 3d). Wang et al.[53] have previously highlighted the key catalytic role that this species plays during growth, viz. accelerating carbon chain growth and facilitating hydrogen transfer reactions. The saturated analogues of this species, e.g. $C_2H_2$ (Figure 3e) and $C_2H_3$ (Figure 3f) have smaller populations by comparison, as expected, considering the catalytic activation of the C-H bonds facilitated by Co discussed above. Our main interest here is to characterise the role of the nanoparticle composition on the populations of these species. In this respect, the evolution of $C_2H$ population during the simulated CVD process is most notable. Figure 3e shows that on pure Co this species is removed from the CVD reaction within 300 ps, ca. 100 ps after carbon supply to the catalyst surface is stopped. However, incorporating Ru into the catalyst nanoparticle increases the lifetime of $C_2H$ on the nanoparticle surface to > 500 ps (beyond the timescale of the simulation employed here). A similar result is observed in Figure 2b,c for $CH_2$ and $CH_3$, respectively, which show that generally the population of $CH_2$ and $CH_3$, respectively, are increased when Ru is present in the nanoparticle, compared to pure Co. From these results it is clear that the catalytic activity of the nanoparticle – i.e. its propensity to activate C-C and C-H bonds – is decreased with increasing Ru content. Results in Figure 3 are therefore consistent with those presented in Figure 2.

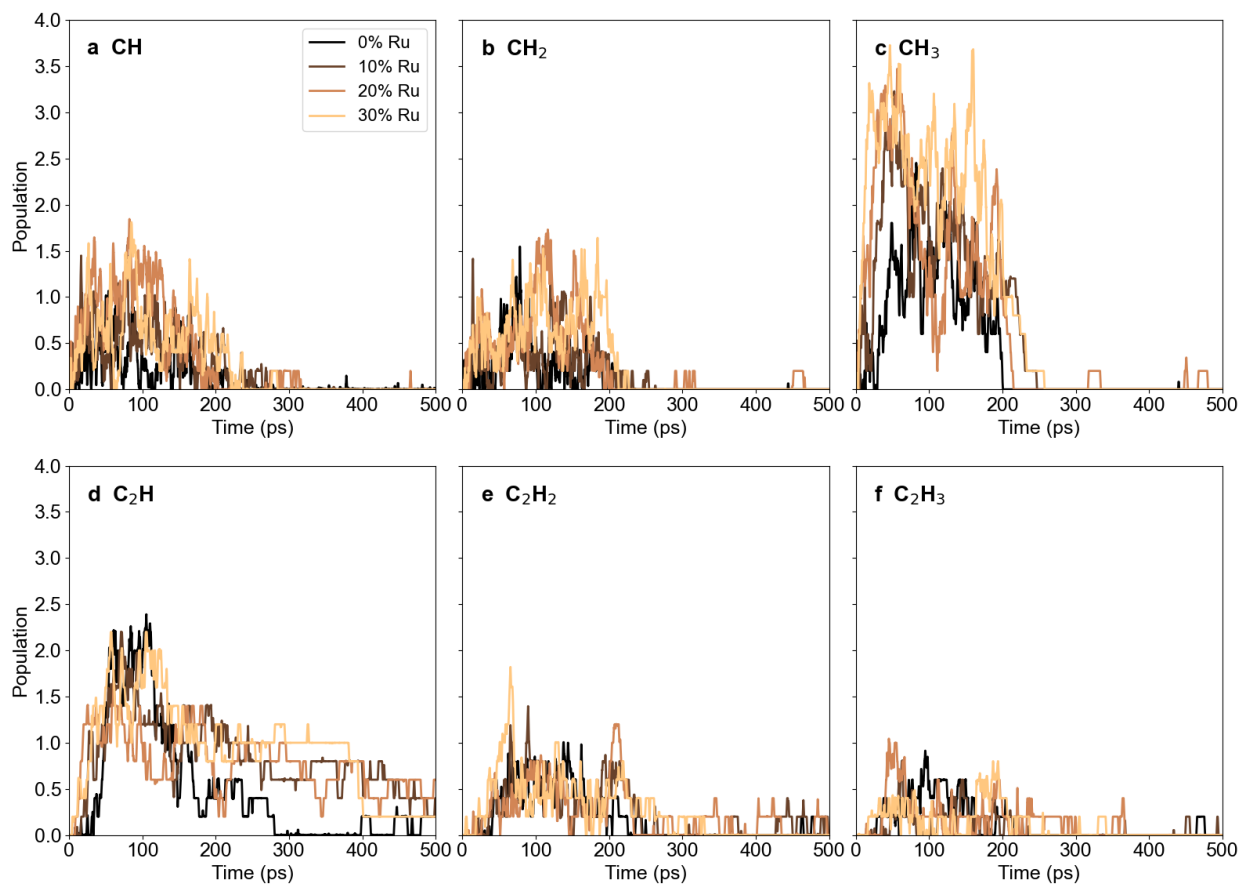

**Figure 3.** Populations of key $C_1$ and $C_2$ SWCNT nucleation precursors observed during simulated CVD on CoRu catalyst nanoparticles as a function of nanoparticle Ru content. (a) CH, (b) $CH_2$, (c) $CH_3$, (d) $C_2H$, (e) $C_2H_2$, (f) $C_2H_3$.

**Carbon Chain and Ring Growth Kinetics**

Following the initial adsorption and activation of the SWCNT growth precursors on the catalyst surface, SWCNT nucleation proceeds via the growth of carbon chains which subsequently oligomerise into $sp^2$-hybridised carbon networks.[6,27,29,53] We consider the impact of Ru content on these two stages of the CVD process in Figure 4 and 5. The results in Figure 4 show that, contrary to trends shown in Figure 2 and 3, the Ru loading in the CoRu nanoparticles has little influence on the chain growth kinetics observed during the early stages of SWCNT nucleation. Here, the evolution of carbon chains on the nanoparticle catalyst surface proceeds in the usual way – $C_2$ first, then $C_3/C_4$, before longer chains become persistent beyond ~300 ps. For $C_n$ chains of essentially all lengths (for *n* up to 11), chain populations are extremely similar for different Ru loadings. This is attributed simply to the fact that, in general, extended carbon chains are adsorbed to the catalyst surface only via a single carbon atom due to hydrogen saturating the remainder of the chain, an effect that is most noticeable for the largest Ru loadings considered here due to the reduced catalytic activity in this case (Figure 4k). This influence of hydrogen in this respect has been noted by us[38,39] and others[53] previously. The single exception to this trend is the population of $C_3$ chains (Figure 4b) which is more persistent on the pure Co catalyst nanoparticle, compared to the CoRu nanoparticles considered here. The rate at which $C_3$ chains are consumed (following their initial formation during 0-150 ps) is also directly proportional to the amount of Ru in the catalyst nanoparticle. One potential explanation for this is the more persistent $C_2H$ population observed when Ru is present in the nanoparticle (discussed above) enables short carbon chains to be converted into longer chains more efficiently. Thus, Ru impacts chain growth kinetics indirectly via its influence on the $C_2H$ population. The relative populations of long carbon chains shown in Figure 4f-i (e.g. for $C_7 - C_{10}$ chains) with and without Ru in the catalyst is consistent with this hypothesis.

The timescale of carbon chain growth is consistent with the timescale upon which the first rings are nucleated (Figure 6b). In this respect the mechanism by which sp-hybridised carbon chains are condensed into $sp^2$-hybridised ring networks is consistent with prior literature, as expected.[6,27,29,53] While no CNT cap was observed on the timescales employed in this work (due to hydrogen saturation), we note that the $sp^2$ networks formed here are pentagon-rich. We therefore expect cap formation to proceed via a mechanism consistent with that of Wang et al., upon hydrogen removal from the edge of the networks.[53] Figure 5 shows that the nanoparticle Ru loading does not yield any clear influence on ring growth kinetics in our simulations. For instance, pentagon growth is slowed in the presence of Ru, but not in a manner proportional to Ru content, while hexagon growth is comparable with and without Ru – except for the Ru 30% loading, which shows a significantly larger population of hexagons within the timescales employed. Figure 5 suggests that the formation of heptagonal defects is increased marginally in the presence of Ru. Both of these results are consistent with the persistence of longer carbon chains in this case (Figure 4f-i), as is the fact that in many instances the ring structures formed are only partially adsorbed to the CoRu nanoparticle surface (Figure 5d).

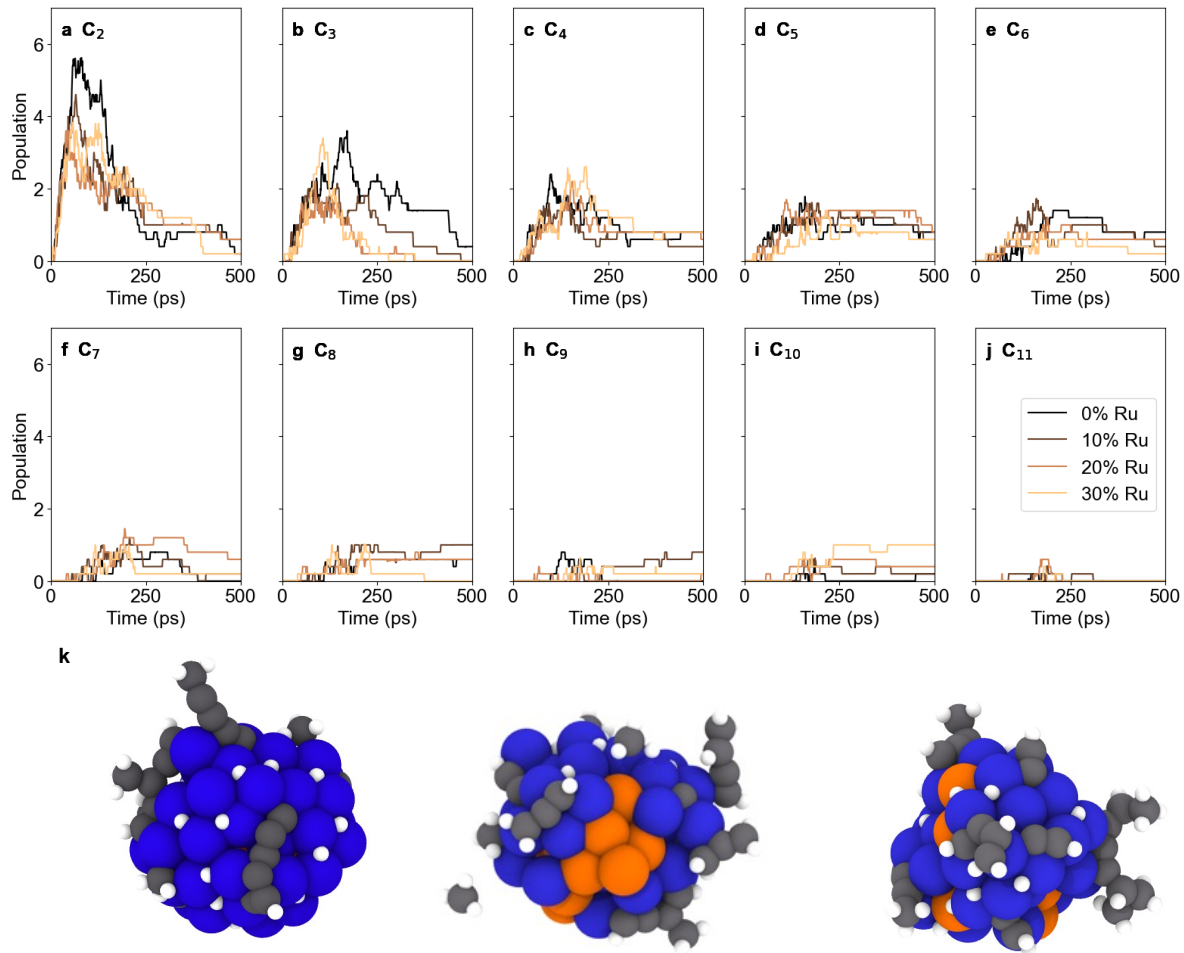

**Figure 4.** (a-j) Carbon chain populations observed during simulated methane CVD on CoRu nanoparticles, as a function of nanoparticle Ru content. (k) Carbon chain adsorption during simulated methane CVD on a CoRu nanoparticle (30% Ru loading) observed in Movie S3 predominantly via lone/terminal carbon atoms due to hydrogen saturation along the chain. Blue, orange, grey and white spheres are Co, Ru, C and H atoms, respectively.

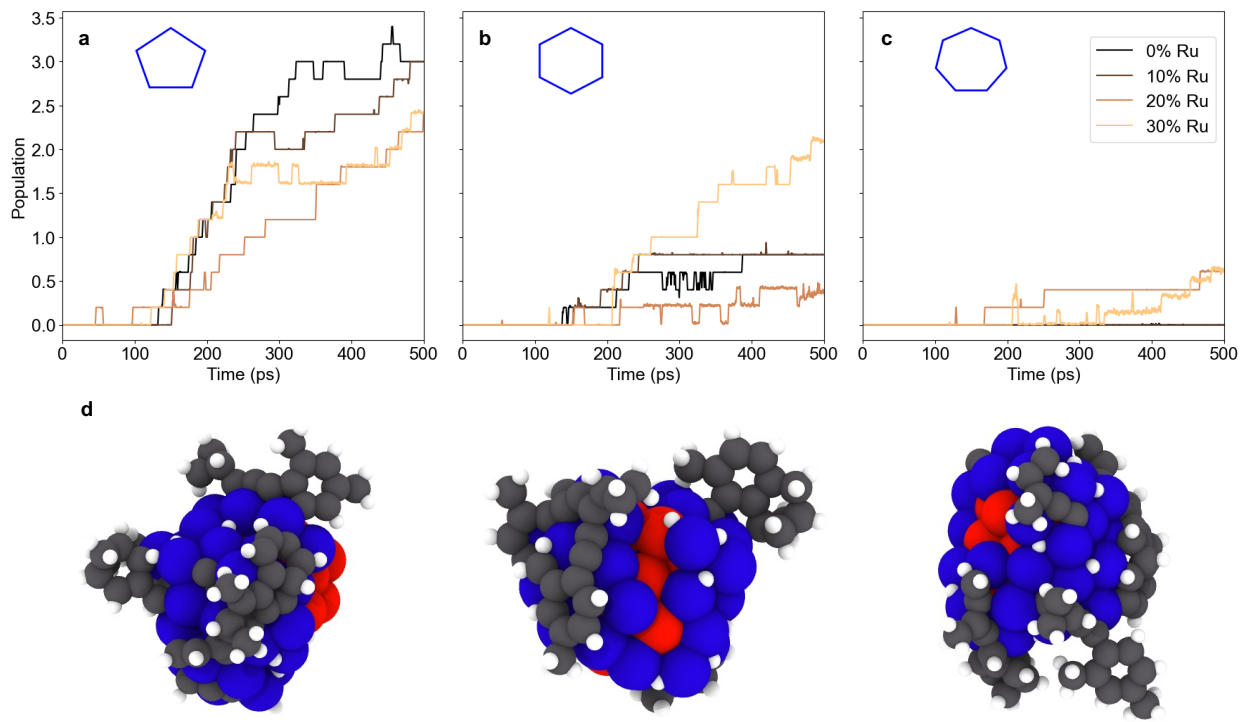

**Figure 5.** (a-c) Carbon ring populations observed during simulated methane CVD on CoRu nanoparticles, as a function of nanoparticle Ru content. (d) Examples of partially adsorbed ring structures observed in Movie S3. Atoms coloured as in Figure 4.

**CoRu Nanoparticle Fermi Level Determines Catalytic Activity**

The results presented above reveal an important insight regarding how the nanoparticle composition influences SWCNT CVD growth. The chemical composition of the nanoparticles considered here is most relevant to the earliest stages of the CVD mechanism itself, viz. precursor adsorption and activation. It is during these elementary stages that our results show Ru plays a direct and influential role in modulating the catalytic activity of the nanoparticle. However, for the subsequent steps in the nucleation and growth process, e.g. chain and ring growth, the nanoparticle Ru content seems less relevant. A simple explanation for this reduction in influence is that these steps in the SWCNT nucleation process do not necessarily occur directly on the nanoparticle surface. Nevertheless, the fact that Ru is essentially absent from the active catalytic interface means that Ru is indirectly steering the catalytic process. Despite not actively participating in the catalytic activation and decomposition, these results are anticipated considering established heuristic models in heterogeneous catalysis, such as d-band theory.[34,54] In the current context, it would be expected that, as Ru content in the catalyst nanoparticle increases, lowering the Fermi level (since Ru 4d states are lower in energy, compared to Co 3d states, Figure 6a). Consequently, the energy difference between the occupied catalyst d states and unoccupied C and H s/p states increases, making C-H bond activation less likely (since this requires electron donation into unoccupied C-H $\sigma^*/\pi^*$ orbitals in surface-adsorbed species). Figure 6b confirms this hypothesis directly; the GFN1-xTB Fermi level of the model CoRu alloy nanoparticles decreases monotonically with the Ru content, across a range of ~0.25 eV.

We propose that invoking established heuristic models such as d-band theory as a new approach to understanding SWCNT CVD growth and a designing CVD alloy catalysts that, to our knowledge, has not been utilised in the SWCNT / graphene growth fields to date. Such models are well-suited for integration into high-throughput screening and machine learning approaches for catalyst discovery,[55] which are underutilised in this field compared to others, such as heterogeneous catalysis. Although our results are based on a small nanoparticle (~1.2 nm in diameter), it is expected that this approach will translate to larger, more experimentally relevant nanoparticle diameters (e.g. > 5-10 nm), since d-band theory itself is based on bulk catalyst phenomena. We note that these results are consistent with our study of FeO-based SWCNT growth catalysts that showed the Fermi level and catalytic efficiency of Fe nanoparticles can be increased by the presence of higher-energy O 2p states.[56]

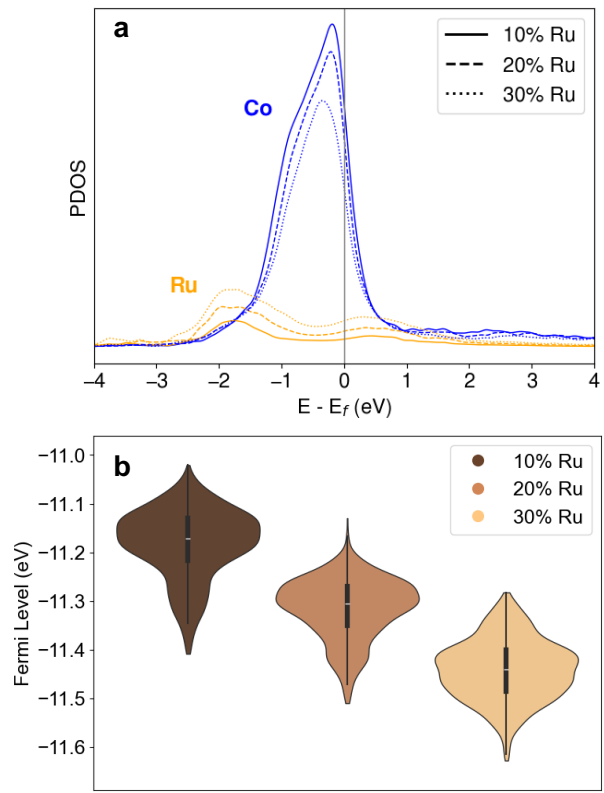

**Figure 6.** Electronic structure of CoRu nanoparticles sampled during simulated CVD as a function of nanoparticle Ru content. (a) Partial density of states (PDOS) of Ru 4p/5s and Co 3d/4s states. (b) CoRu nanoparticle Fermi level distributions. According to d-band theory, the CoRu nanoparticle's catalytic activity is expected to decrease with Ru content as the valence d-band centre shifts away from the (constant) energy of unoccupied C/H states in the conduction band.

## Conclusions

We have presented quantum chemical molecular dynamics simulations of methane CVD on model CoRu nanoparticles with varied Ru content. Our results reveal that Ru influences this catalytic process indirectly in a manner that is consistent with established heuristic models for heterogeneous catalysis, such as d-band theory.[34,54] By decreasing the energy of the nanoparticle Fermi level, the Ru component of these alloy nanoparticles decreases the catalytic efficiency of the active Co component of the nanoparticle. In turn, C-H bond activation is impeded, key reactive intermediate species (e.g. $C_2H$) become longer-lived, and longer carbon chains are stabilised through the earliest stages of SWCNT nucleation. These phenomena also impact the formation of carbon rings, with hexagon rings being notably more likely for the highest Ru loadings considered here. Our results show that CoRu nanoparticles at CVD-relevant conditions likely exist predominantly as a Ru-Co core-shell structure, but this structure is highly dynamic and influenced strongly by the adsorption of carbon during the CVD process itself. The fact that Ru influences the catalytic activity of the nanoparticle overall without actively participating in the catalytic process itself means that SWCNT growth catalyst composition, is an important determinant of its catalytic activity and a key factor that should be considered in designing intermetallic catalysts for SWCNT growth.

**Supporting Information**

Movies of simulated methane CVD on CoRu nanoparticles for each Ru content; comparison of GFN1-xTB and PBE energies and forces observed during simulated CVD on CoRu nanoparticles; radial distribution functions of CoRu nanoparticles as a function of temperature (300 – 2100 K); atomic Lindemann indices for the Co and Ru components of CoRu nanoparticles between 300 – 2100 K; populations of $C_xH_y$ (x = 1-4) nucleation precursors observed during simulated CVD on CoRu nanoparticles as a function of Ru content.

**Acknowledgements**

AJP acknowledges Australian Research Council funding (ARC DP210100873). The authors thank Dr Benji Maruyama (Air Force Research Labs) for helpful discussions. This research was undertaken with the assistance of resources provided at the NCI National Facility systems at the Australian National University, through the National Computational Merit Allocation Scheme supported by the Australian Government.

# Electronic Structure of Bimetallic CoRu Catalysts Modulates SWCNT Nucleation

Alister J. Page*,[1] Dan Villamanca,[1] Placidus B. Amama,[2] Ben McLean[3]

[1]*Discipline of Chemistry, College of Engineering, Science and Environment, The University of Newcastle, Callaghan NSW 2308, Australia*

[2] *Tim Taylor Department of Chemical Engineering, Kansas State University, Manhattan, KS 66506, USA*

[3]*School of Engineering, RMIT University, Victoria, 3001, Australia*

*Corresponding Author.  Email: alister.page@newcastle.edu.au

## *Supporting Information*

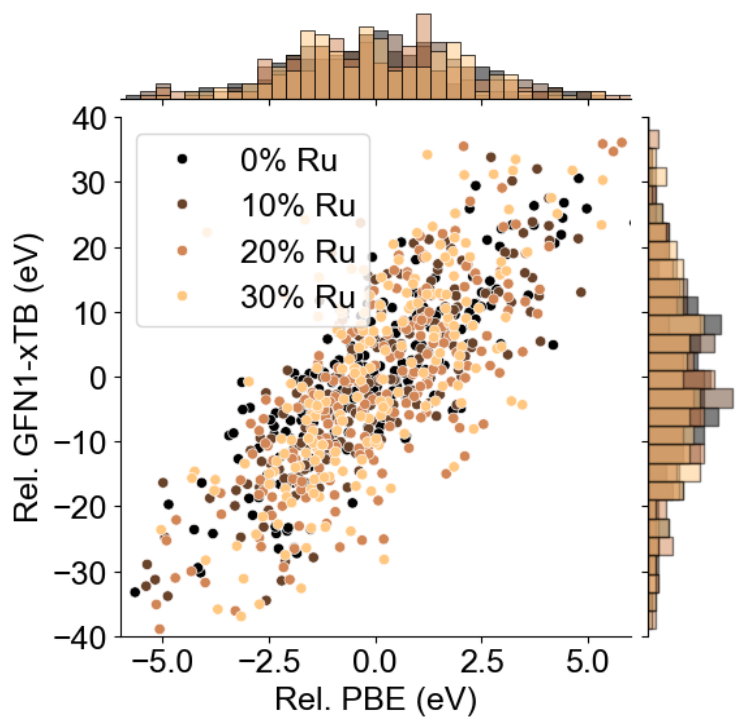

**Figure S1.** Comparison of GFN1-xTB and PBE energies (eV) observed during NVT-MD simulations of CoRu nanoparticles as a function of Ru content. Energies are scaled relative to the mean energy value in each case for direct comparison.

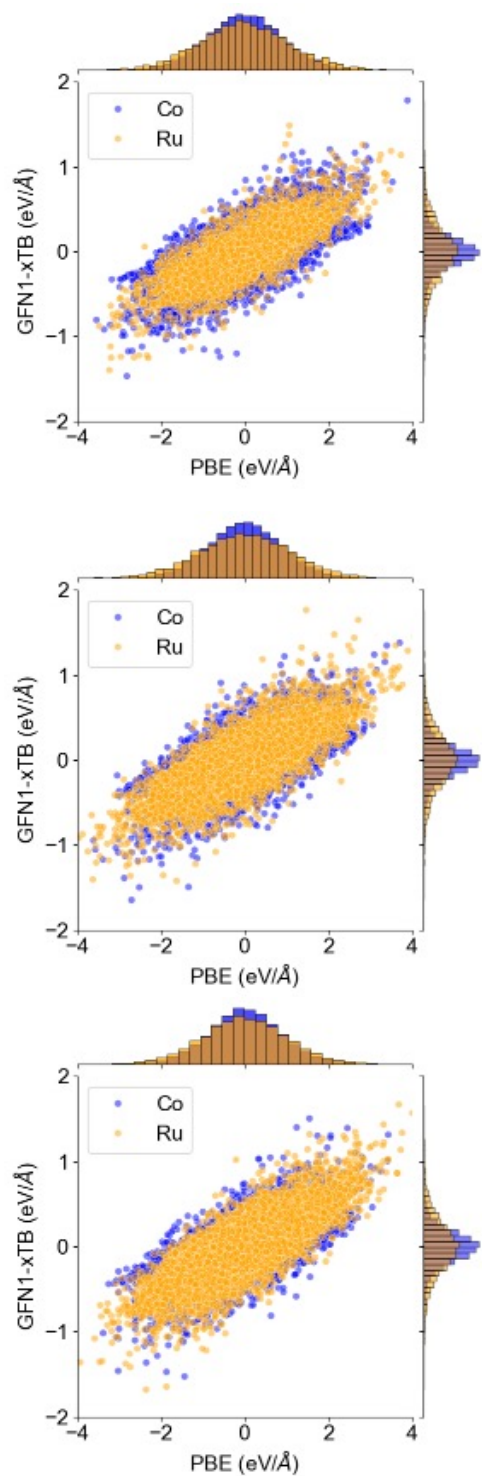

**Figure S2.** Comparison of GFN1-xTB and PBE force components (eV/Å) observed during NVT-MD simulations of CoRu nanoparticles as a function of Ru content (10% top; 20% middle; 30% bottom).

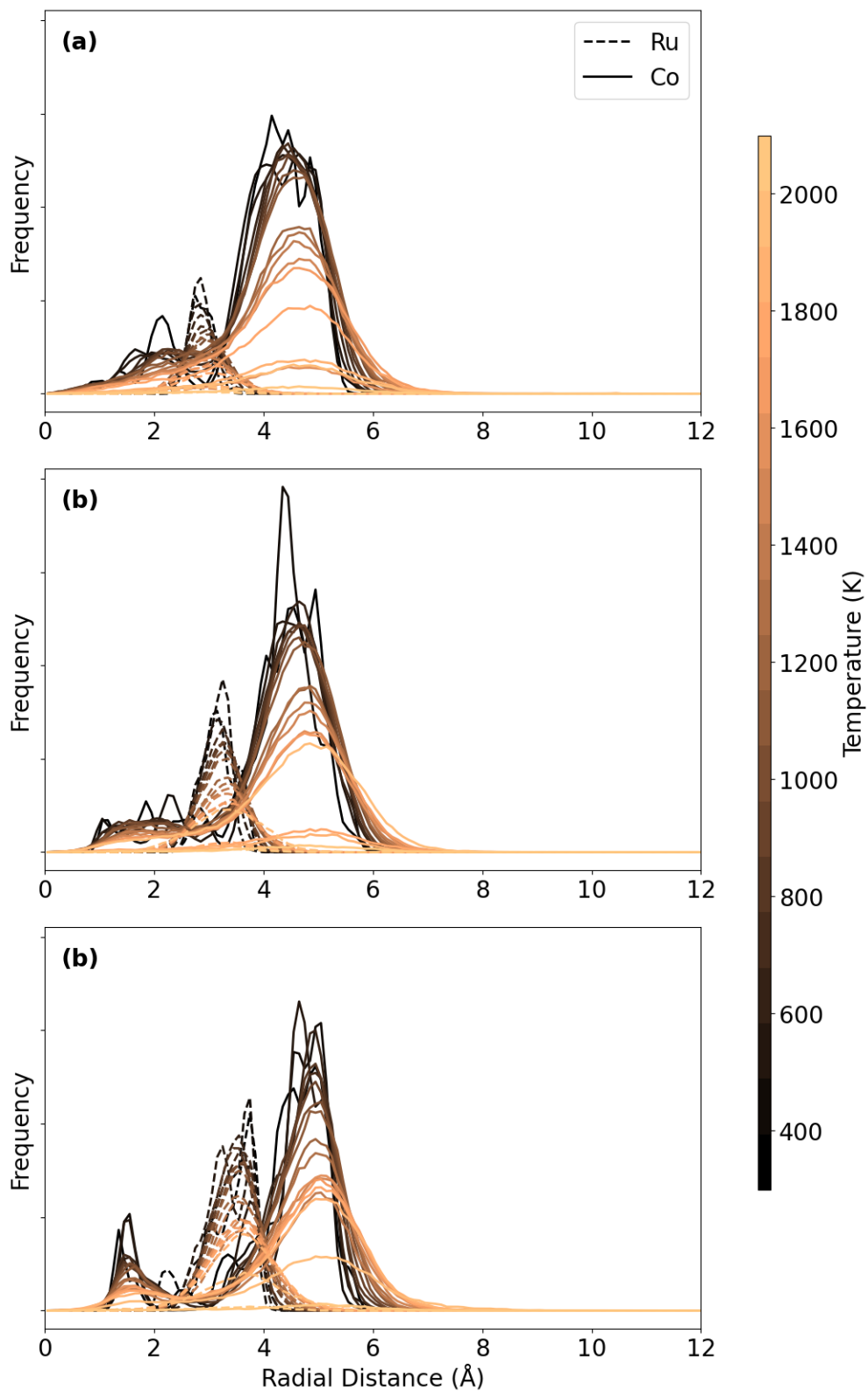

**Figure S3.** Temperature dependence of radial distributions of Co and Ru atoms in the CoRu alloy nanoparticle between temperatures of 298K and 2000K. Distribution functions are based on 50 ps of GFN1-xTB/MD simulations at each respective temperature, in the absence of carbon.

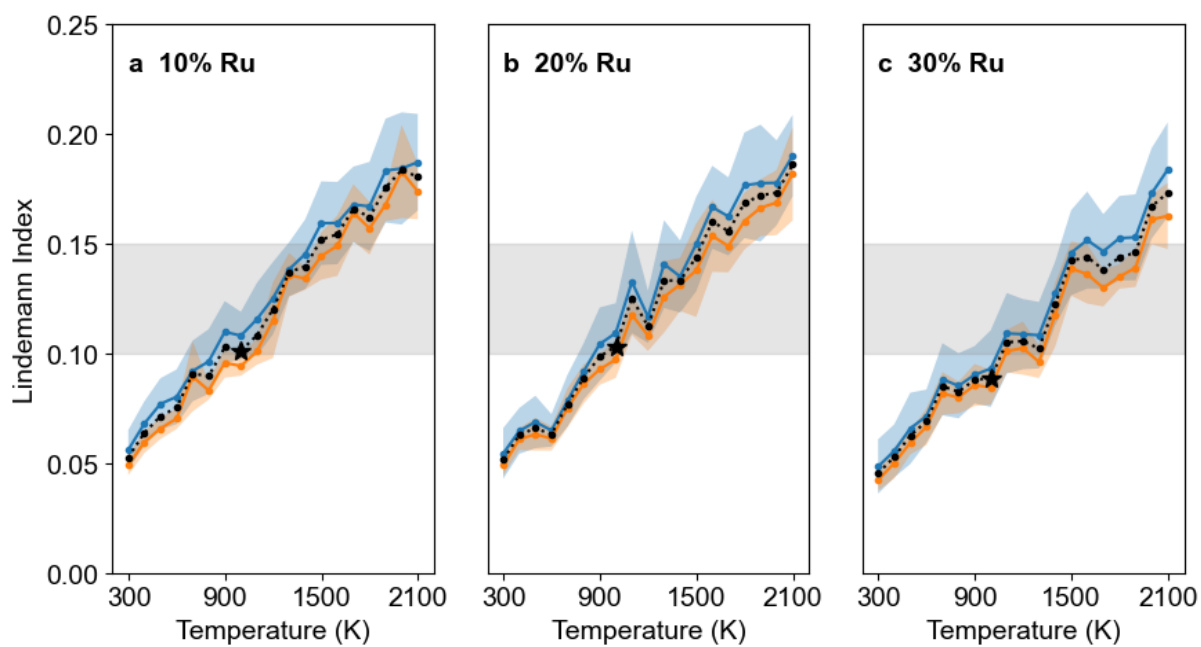

**Figure S4.** Lindemann indices for the Co and Ru atoms in the CoRu nanoparticles considered here for (a) 10 % Ru, (b) 20% Ru and (c) 30% Ru. The dotted black line indicates the nanoparticle average, and the shaded regions indicate ±1 standard deviation. The shaded grey region indicates the commonly accepted range in the Lindemann index that signifies solid-liquid phase transition, and the black stars indicate the nanoparticle Lindemann index at the CVD simulation temperature of 1000 K.

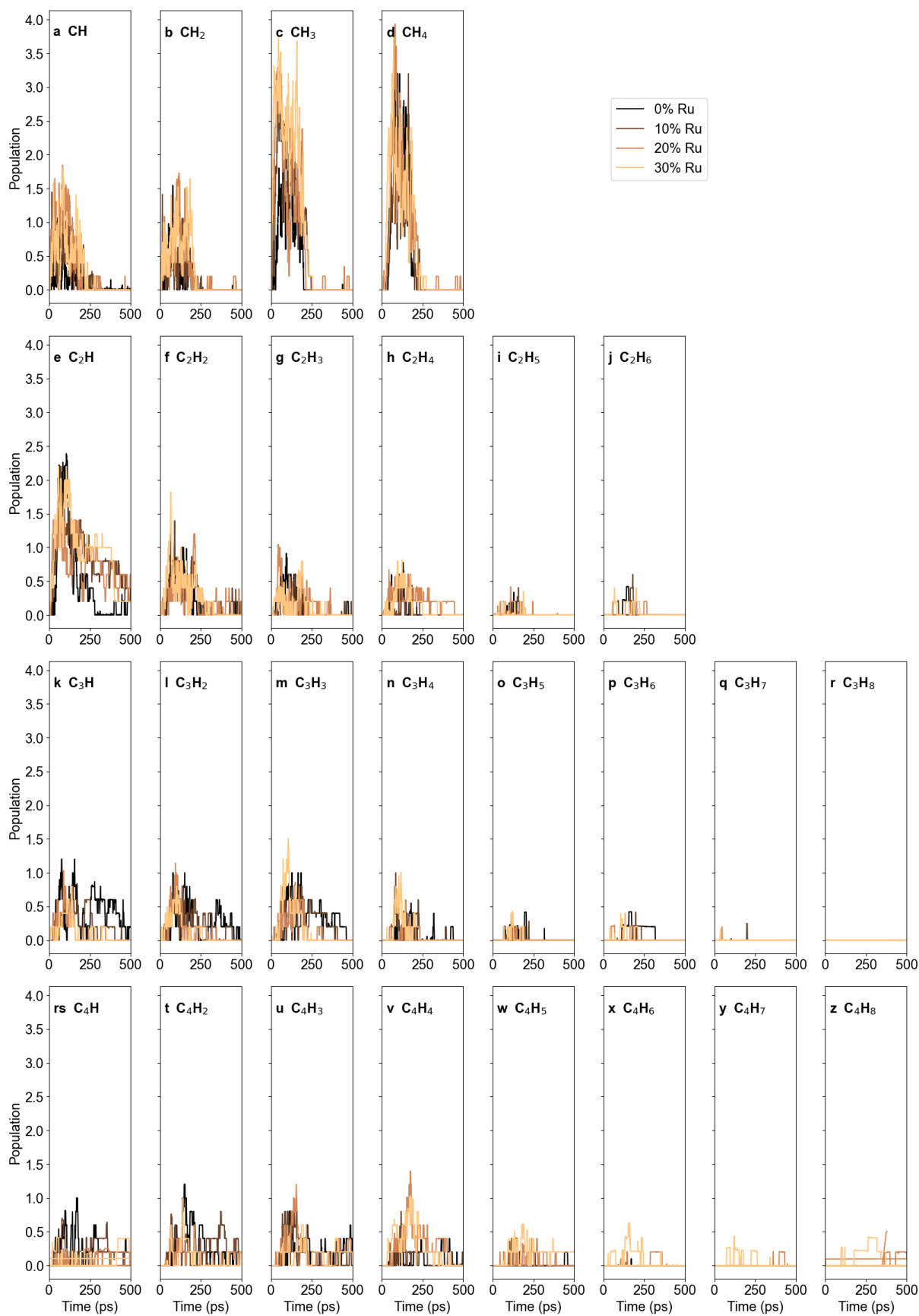

**Figure S5.** Populations of $C_x$ nucleation precursors ($x$ = 1-4) observed during simulated CVD on CoRu nanoparticles as a function of nanoparticle Ru content.